\documentclass[twocolumn,showpacs,aps,preprintnumbers,amsmath,amssymb,10pt]{revtex4-1}
\usepackage{color}
\usepackage{hyperref}
\usepackage{graphicx}
\usepackage{dcolumn}
\usepackage{bm}
\usepackage{epstopdf}
\usepackage{braket}

\begin{document}

\title{Dynamics of nuclear spin polarization induced and detected by coherently precessing electron spins in fluorine-doped ZnSe}

\author{F.~Heisterkamp$^{1}$, E.~Kirstein$^{1}$ A.~Greilich$^{1}$, E.~A.~Zhukov$^{1}$}
\author{T.~Kazimierczuk$^{1}$}
\thanks{Present address: Institute of Experimental Physics, Faculty of
Physics, University of Warsaw, 02-093 Warsaw, Poland.}
\author{D.~R.~Yakovlev$^{1,2}$, A.~Pawlis$^{3}$, and M.~Bayer$^{1,2}$}
\affiliation{$^1$ Experimentelle Physik 2, Technische Universit\"at
Dortmund, 44221 Dortmund, Germany}
\affiliation{$^2$ Ioffe Institute, Russian Academy of Sciences, 194021 St.~Petersburg, Russia}
\affiliation{$^3$ Peter Gr\"unberg Institute (PGI-9), Forschungszentrum J\"ulich, 52425 J\"ulich, Germany}

\date{\today}

\begin{abstract}
We study the dynamics of optically-induced nuclear spin polarization
in a fluorine-doped ZnSe epilayer via time-resolved Kerr rotation.
The nuclear polarization in the vicinity of a fluorine donor is
induced by interaction with coherently precessing electron spins in
a magnetic field applied in the Voigt geometry. It is detected by
nuclei-induced changes in the electron spin coherence signal. This
all-optical technique allows us to measure the longitudinal spin
relaxation time $T_{1}$ of the $^{77}\text{Se}$ isotope in a
magnetic field range from 10 to 130~mT under illumination. We
combine the optical technique with radio frequency methods to
address the coherent spin dynamics of the nuclei and measure Rabi
oscillations, Ramsey fringes and the nuclear spin echo. The
inhomogeneous spin dephasing time $T_{2}^{*}$ and the spin coherence
time $T_{2}$ of the $^{77}\text{Se}$ isotope are measured. While the
$T_{1}$ time is on the order of several milliseconds, the $T_{2}$
time is several hundred microseconds. The experimentally determined
condition $T_{1}\gg T_{2}$ verifies the validity of the classical
model of nuclear spin cooling for describing the optically-induced
nuclear spin polarization.
\end{abstract}


\maketitle 

The spin of a donor-bound electron in fluorine-doped ZnSe represents
a promising system for quantum information technologies. So far
emission of indistinguishable single photons~\cite{Sanaka09}, photon
entanglement~\cite{SanakaNL12} and optical control of single
electron spins~\cite{DeGreve10,Sanaka12,SleiterNL13} were achieved.
Studies of the electron spin dynamics of an ensemble of donor-bound
electrons demonstrated a spin dephasing time $T_{2}^{*}$ of
$33\,\text{ns}$ and a longitudinal spin relaxation time $T_{1}$ of
$1.6\,\mu\text{s}$~\cite{Greilich12,Heisterkamp06_2015}. At low
temperatures the hyperfine interaction with nuclear spins is the
main mechanism limiting the electron spin coherence. ZnSe is a
particularly attractive material due to the low natural abundance of
isotopes with nonzero nuclear spins ($4.11\%$ $^{67}$Zn ($I=5/2$)
and $7.58\%$ $^{77}$Se ($I=1/2$)) and the possibility to purify it
isotopically such that it does not contain nonzero nuclear spins.
However, this approach is technologically demanding. Alternatively,
one may search for effects where the polarization of nuclear spins
provides favorable conditions for a long-lived electron spin
coherence. For example, the nuclear frequency focusing effect of the
electron spin coherence in singly-charged $(\text{In,Ga})\text{As}/\text{GaAs}$ quantum
dots reduces the electron spin dephasing by driving the precessing
ensemble towards a single mode collective motion~\cite{Greilich2007}.
In order to understand the underlying mechanisms comprehensive
information on the polarization and relaxation dynamics of the
nuclei interacting with the electrons driven by periodic laser
excitation is required.

We recently demonstrated a spatially inhomogeneous nuclear spin
polarization induced in the vicinity of the fluorine donor in a
picosecond pump-probe Kerr rotation (KR) experiment on a fluorine-doped
ZnSe epilayer~\cite{Heisterkamp_12_2015}. The nuclear spin
polarization occurs under excitation with a helicity modulated pump
beam, for which the induced average electron spin polarization is
expected to be zero. The classical model of nuclear spin cooling was
used to explain the induced nuclear spin
polarization~\cite{OO_Chapter5}. It implies that the nuclear spin
system can be described using the spin temperature approach, which
requires that the longitudinal spin relaxation time $T_1$ is much
longer than the spin coherence time $T_2$: $T_{1}\gg T_{2}$. To
validate this condition these nuclear relaxation times need to be
measured under the conditions of the pump-probe experiment. 

The feasibility of such measurements has been demonstrated in Refs.~\cite{Sanada2006,Kondo2008} on $\text{GaAs}/(\text{Al,Ga})\text{As}$ QWs under optical excitation
with a constant circular polarization.  Similar to these studies we
combine the time-resolved Kerr rotation (TRKR) measurements with radio frequency (RF) techniques to study the
coherent spin dynamics of the nuclei, but we perform our studies
under helicity modulated excitation. Furthermore, we present an all-optical technique (see also
Ref.~\cite{Zhu2014}) employing TRKR in the resonant spin amplification (RSA)
regime~\cite{Kikkawa98} to also measure the $T_{1}$ time. We use these two methods to perform a complete study of the dynamics of the $^{77}\text{Se}$ isotope, whose polarization was
studied in Ref.~\cite{Heisterkamp_12_2015}, under the conditions of the TRKR experiment and measure the nuclear spin relaxation time $T_{1}$, the
inhomogeneous spin dephasing time $T_{2}^{*}$ and the spin coherence
time $T_{2}$. From the results we conclude that the spin temperature approach is valid under these experimental conditions.

We employ a pump-probe scheme to measure TRKR using a mode-locked
Ti:Sapphire laser with a pulse duration of $1.5\,\text{ps}$ at a
repetition rate of $75.75\,\text{MHz}$. The circularly polarized
pump beam excites the sample along the growth axis and the KR of the linearly polarized probe beam is measured with
a balanced photoreceiver connected to a lock-in amplifier. The pump
beam is helicity-modulated by an electro-optical modulator with the
frequency $f_{\text{m}}$ varied from 50 up to 1050~kHz. We conduct
all measurements at a fixed small negative time delay of the probe
pulses with respect to the pump pulses and scan the magnetic field,
i.e., use the RSA regime~\cite{Kikkawa98,Yugova12}. The probe beam is
kept unmodulated. Both beams have a diameter of about
$300\,\mu\text{m}$ on the sample. The pump power is kept constant at
$8\,\text{mW}$ and the probe power at $0.5\,\text{mW}$ for all
measurements. 

We measure the nuclear spin polarization by its influence on the Larmor precession 
frequency of the donor-bound electron spins in a fluorine-doped ZnSe epilayer with 
a dopant concentration of about $10^{18}\,\text{cm}^{-3}$. The sample was
grown by molecular-beam epitaxy on $(001)$-oriented GaAs substrate.
For details about the  optical properties and electron spin
coherence in this sample we refer to Ref.~\cite{Greilich12}
(Sample C). We use a degenerate pump-probe scheme and resonantly
excite the donor-bound heavy hole exciton
($\text{D}^{0}\text{X}-\text{HH}$) at $2.800\,\text{eV}$. To obtain
the required photon energy the laser photon energy is doubled by a
beta barium borate (BBO) crystal. The sample is placed in an optical
cryostat with a superconducting split coil solenoid with the
magnetic field oriented perpendicular to the optical axis and the
structure growth axis. The sample temperature is fixed at about
$1.8\,\text{K}$.

A RF coil near the sample surface allows us to
apply in addition RF fields with variable frequency from $50$ to
$500\,\text{kHz}$. The axis of the coil is oriented along the
optical axis. Thus, the oscillating magnetic field is oriented along
the optical axis and perpendicular to the external magnetic field. A
delay generator allows us to apply well-defined and exactly timed
sequences of RF pulses to perform experiments on the nuclear spin
coherence.

For the $T_{1}$ measurements we use a TTL multiplexer (average
propagation delay: $12\,\text{ns}$) in combination with two
arbitrary function generators to quickly switch $f_{\text{m}}$ (with
a rise time of 8 ns). The signal is continuously demodulated
by two lock-in amplifiers, each locked permanently on one of the two
modulation frequencies. Their signals are recorded by a fast
digitizer card. The time resolution of this setup is given by the
digitizer card and the lock-in amplifier, which is the limiting
factor here and demodulates the signal at a time constant of
$50\,\mu\text{s}$.

The nuclear spin relaxation time $T_{1}$ is measured using an
all-optical approach based on fast switching between two different
modulation frequencies $f_{\text{m},1}$ and $f_{\text{m},2}$. The
first modulation frequency $f_{\text{m},1}$ is close to the
optically-induced nuclear magnetic resonance (NMR) at the particular magnetic field, while the
second one is set to $f_{\text{m},2}=1050\,\text{kHz}$, where the
NMR can be reached only at a stronger magnetic field.
\begin{figure}[tb]
\includegraphics[width=1.0\columnwidth]{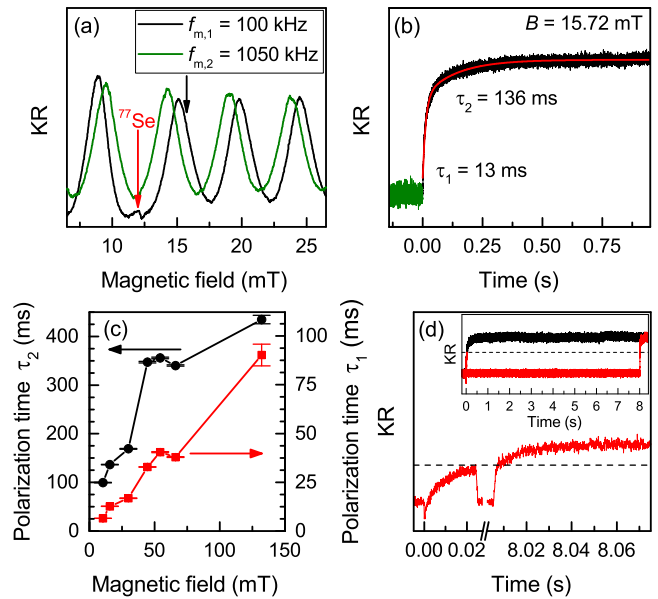}
\caption{(Color online) (a) RSA spectra measured at
$f_{\text{m},1}=100$\,kHz (black line) and
$f_{\text{m},2}=1050$\,kHz (green line). The red arrow marks the
optically induced NMR of the $^{77}\text{Se}$ isotope at
$f_{\text{m},1}=100\,\text{kHz}$ and the black arrow marks the
magnetic field position ($B=15.72\,\text{mT}$) for the measurement
shown in Fig.~\ref{fig:T1}(b). (b) Change of KR amplitude at fixed
magnetic field induced by switching from $f_{\text{m},2}$ (green
line) to $f_{\text{m},1}$ (black line). Red line shows a double
exponential fit to the data. (c) Polarization times in dependence on
the magnetic field strength. Lines are shown as guides to the eye.
(d) Change of KR signal (black line) at $B=5.62\,\text{mT}$ induced
by switching from $f_{\text{m},1}=50\,\text{kHz}$ to
$f_{\text{m},2}=1050\,\text{kHz}$. Red line shows a measurement at
the same conditions but with a dark time of $8\,\text{s}$. Dashed black line at the KR amplitude upon closing the shutter is
shown as a guide to the eye. Inset shows the same signals, but
over the full time range of about $8\,\text{s}$.} \label{fig:T1}
\end{figure}

Figure~\ref{fig:T1}(a) shows two RSA spectra measured at modulation
frequencies of $f_{\text{m},1}=100\,\text{kHz}$ (black line) and
$f_{\text{m},2}=1050\,\text{kHz}$ (green line), respectively. The
magnetic field is varied which, in turn, leads to a change of the
electron spin Larmor precession frequency, so that the spins precess
either in phase or out of phase with the laser repetition frequency
and the RSA signal exhibits characteristic periodic peaks in
dependence on the magnetic field perpendicular to the optical axis
(Voigt geometry)~\cite{Yugova12}. Due to optically-induced,
inhomogeneous nuclear polarization (see
Ref.~\cite{Heisterkamp_12_2015} for details) the effective
field which determines the electron spin precession depends on the
modulation frequency. Thus, the RSA peaks at different modulation
frequencies are shifted relative to each other. The red arrow marks
the position of the optically induced NMR of the $^{77}$Se isotope
at $f_{\text{m},1}=100\,\text{kHz}$. The black arrow marks the
magnetic field position ($B=15.72\,\text{mT}$) for the measurement
of the $T_1$ time. Figure~\ref{fig:T1}(b) illustrates the change of
the KR signal when the modulation frequency is switched from
$f_{\text{m},2}=1050\,\text{kHz}$ to
$f_{\text{m},1}=100\,\text{kHz}$. The green line represents the KR
signal at $f_{\text{m},2}=1050\,\text{kHz}$. The black line shows
the transient of the KR signal after switching to
$f_{\text{m},1}=100\,\text{kHz}$.  By measuring the KR signal at a
fixed magnetic field one detects the shift of the RSA peak from its
position at $f_{\text{m},2}$ to its position at $f_{\text{m},1}$.
The KR signal increases and saturates in less than a second. The
rise of the KR signal is fitted with a double exponential function yielding
rise times of $\tau_{1}=13\pm1\,\text{ms}$ and
$\tau_{2}=136\pm1\,\text{ms}$. We interpret these components as the
minimal and maximal polarization time of the repolarization process
with a stretched exponent. We tentatively assign the fastest
polarization time to the strongly polarized nuclei near the center
of the donors which are exposed to the strongest Knight field and
are most sensitive to a change of modulation frequency. On the other
hand, the longest polarization time should result from the weaker
polarized nuclei located farthest from the donors. These nuclei
interact with a much weaker Knight field, so here the repolarization
process occurs at a longer timescale.

Figure~\ref{fig:T1}(c) shows these polarization times in dependence on
the magnetic field. For these measurements $f_{\text{m},1}$ is
adjusted correspondingly to stay close to the optically-induced NMR
at higher fields.  The red squares represent the minimal
polarization time (right axis), while the black circles shows the
maximal polarization time (left axis). Both components increase with
magnetic field, which we tentatively assign to the increasing
difference of the electron and the nuclear Zeeman splitting. We
conclude that the time to polarize the nuclei is on the order of
several tens of milliseconds (fastest) or several hundreds of
milliseconds (longest) in the magnetic field range from $10$ to
$130\,\text{mT}$.

Note that all these measurements are conducted under illumination. A lower limit
for the nuclear spin relaxation time in
darkness without illumination $T_{1}^{\text{dark}}$ can be estimated from the measurement
shown in Fig.~\ref{fig:T1}(d). Here the switching of $f_{\text{m}}$ is
combined with a shutter, which simultaneously blocks the pump and the
probe beam. While the black line shows the continuous transient of
the KR signal upon switching from $f_{\text{m},1}=50\,\text{kHz}$ to
$f_{\text{m},2}=1050\,\text{kHz}$ modulation at $B=5.62\,\text{mT}$,
the red line is measured with an additional dark time of
$8\,\text{s}$. The shutter is closed at about $24\,\text{ms}$ after
the modulation frequency has been switched and reopened at about
$8.006\,\text{s}$. Closing the shutter slows down the nuclear spin
relaxation process, since it prevents a repolarization or
depolarization due to spin flip-flops with spin polarized
electrons~\cite{Korenev2007}. The amplitudes of the KR signals
before and after this dark time are nearly the same. Thus, we
conclude that $T_{1}^{\text{dark}}$ exceeds several tens of seconds.
\begin{figure}[tb]
\includegraphics[width=\columnwidth]{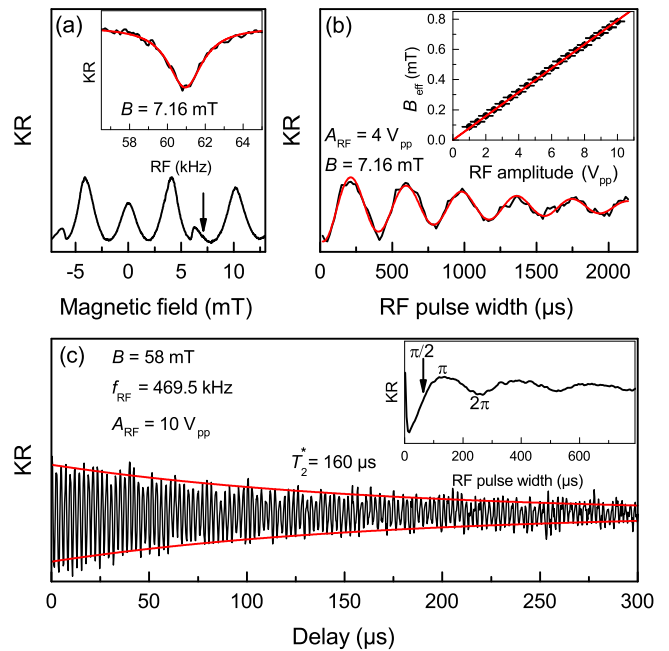}
\caption{(Color online) (a) RSA spectrum measured at
$f_{\text{m}}=50$\,kHz. The arrow shows the magnetic field position
for the measurement of the KR signal in dependence on the applied RF
fields. The inset shows the change of the KR signal induced by an
RF-field of 0.5\,V$_{\text{pp}}$. (b) Rabi oscillations of $^{77}$Se
measured with 4\,V$_{\text{pp}}$ at $f_{\text{RF}}=60.9\,\text{kHz}$
as a function of the RF pulse width. The inset shows the effective
magnetic field of the RF coil in dependence on the RF voltage. (c)
Ramsey fringes with a period of $2.33\mu\text{s}$. The fit with an exponentially damped oscillation
is shown by its envelope (red line). The inset shows a Rabi
oscillations measurement used to determine the length of a $\pi/2$
pulse.} \label{fig:T2star}
\end{figure}

Now we turn to measurements based on coherent manipulation of the
nuclear spins with RF fields. The small peaks in the RSA spectrum at
about $B=\pm6.3\,\text{mT}$, shown in Fig.~\ref{fig:T2star}(a), are
caused by the NMR induced by helicity modulation of the pump with
$f_{\text{m}}=50\,\text{kHz}$. In order to coherently control the
nuclear spins one needs to determine the nuclear magnetic resonance frequency
$f_{\text{NMR}}$ at fixed external magnetic field. The inset shows
an RF sweep with a peak-to-peak amplitude of $0.5\,\text{V}$
($A_{\text{RF}}=0.5\,\text{V}_{\text{pp}}$) to determine
$f_{\text{NMR}}$. The measurement is performed at $B=7.16\,\text{mT}$ (see arrow), where the KR
signal is sensitive to the RF excitation of the nuclei. This RF
excitation depolarizes the nuclei and thereby reduces the Overhauser
field component along the external field ($B_{\text{N}}\parallel
B$), if its frequency is at or close to $f_{\text{NMR}}$ in the
external magnetic field, so that the optically-induced NMR vanishes
and the KR signal exhibits a dip around $f_{\text{NMR}}$. The red
line is a Lorentzian fit used to determine the NMR frequency
$f_{\text{NMR}}=60.93\pm0.01\,\text{kHz}$.

We switch from this continuous-wave (CW) RF excitation to RF pulses
of well-defined width to investigate the coherent properties of the
nuclear spins. Figure~\ref{fig:T2star}(b) shows how the amplitude of
the measured KR signal depends on the width of the RF pulses at
resonant excitation of the NMR. We observe oscillations, which we
interpret as Rabi oscillations, caused by the rotation of the
nuclear spins around the effective magnetic field produced by the RF
coil~\cite{Rabi1937,Rabi1938,Rabi1939}. The red line is a fit with an
exponentially damped oscillation which yields the Rabi frequency
$f_{\text{R}}=2.6\,\text{kHz}$. This frequency can be used to
calculate the effective induced magnetic field at a given RF voltage
using the relation $B_{\text{eff}}[\text{mT}]=
0.1226\,[\text{mT}/\text{kHz}] \cdot f_{\text{R}}[\text{kHz}]$,
where the gyromagnetic ratio $\gamma$ for the $^{77}\text{Se}$
isotope $\gamma=5.1253857\cdot10^{7}\,\text{rad}\text{s}^{-1}\text{T}^{-1}$
(see Ref.~\cite{Harris2001}) is used in $B_{\text{eff}}=2\pi
f_{\text{R}}/\gamma$. For the curve in Fig.~\ref{fig:T2star}(b) this
yields $B_{\text{eff}}=320\,\mu\text{T}$. The inset of
Fig.~\ref{fig:T2star}(b) shows the effective magnetic field in
dependence on the RF amplitude. We obtain
$B_{\text{eff}}\,[\mu\text{T}]=79\,[\mu\text{T}/\text{V}_{\text{pp}}]\cdot
A_{\text{RF}}[\text{V}_{\text{pp}}]$.

The inset in Fig.~\ref{fig:T2star}(c) shows a Rabi oscillation
measurement at $B=58\,\text{mT}$, $f_{\text{RF}}=429.5\,\text{kHz}$
and $A_{\text{RF}}=10\,\text{V}_{\text{pp}}$. It is used to
determine the RF pulse width of a $\pi/2$ pulse. Using two $\pi/2$
pulses with a controllable delay $\tau$ between them allows one to
measure Ramsey fringes and thereby determine the inhomogeneous spin
dephasing time $T_{2}^{*}$ of the nuclear
spins~\cite{Ramsey1949,Ramsey1950,Ramsey1995}.
Figure~\ref{fig:T2star}(c) demonstrates such a measurement. The
first $\pi/2$ pulse creates a coherent superposition of the nuclear
spins between the ground state $\ket{0}$ and the excited state
$\ket{1}$ (both defined with respect to the constant external
magnetic field) on the equator of the Bloch sphere~\cite{Bloch1946}.
The spins then precess in the equatorial plane, whereat the
precession frequency is given by the Zeeman splitting of the nuclear
spins. Due to this precession the nuclear spins acquire a relative
phase with respect to the second $\pi/2$ pulse, so that this pulse
will rotate the spins either to the $\ket{0}$ or the $\ket{1}$
state. The KR signal as a function of the delay $\tau$, in turn,
displays oscillations due to this periodic change between $\ket{0}$
and $\ket{1}$. The red line in Fig.~\ref{fig:T2star}(c) shows the
envelope of a fit with an exponentially damped oscillation which
yields $T_{2}^{*}=160 \pm 5\,\mu\text{s}$. At this time the nuclei
run out of phase in their coherent precession. 

\begin{figure}[tb]
\includegraphics[width=\columnwidth]{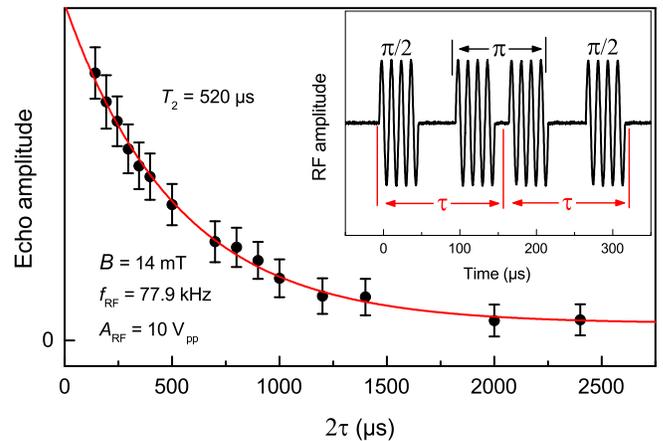}
\caption{Nuclear spin echo amplitude in dependence on the total time
delay $2\tau$. The red line is an exponential fit. The inset
illustrates the employed RF pulse scheme (time dependence of the
normalized RF amplitude).} \label{fig:T2}
\end{figure}

To determine the nuclear spin coherence time $T_{2}$ one needs to
apply an additional $\pi$ pulse in between the two $\pi/2$ pulses
used for the Ramsey method. The $\pi$ pulse applied after a time
$\tau$ inverts the orientation of the spins which then rephase
during the subsequent, second interval $\tau$. This leads to a Hahn
echo (nuclear spin echo)~\cite{Hahn1950} at the time $2\tau$ when
the dephasing of the spins due to ensemble inhomogeneities is
completely compensated due to inversion of the system at $\tau$. The
decay of the echo amplitude which is determined at approximately
$2\tau$ yields the $T_2$ time. Figure~\ref{fig:T2} demonstrates such
a measurement at $B=14\,\text{mT}$ and
$A_{\text{RF}}=10\,\text{V}_{\text{pp}}$. The echo amplitude in
dependence on the total time delay ($2\tau$) between the $\pi/2$
pulses is best fitted with an exponential decay. This fit yields
$T_{2}=520\pm 25\,\mu\text{s}$. The inset shows the used pulse
sequence . The $\pi$ pulse is constructed of two $\pi/2$
pulses of approximately $50\,\mu\text{s}$ length (4 cycles at
$f_{\text{RF}}=77.9\,\text{kHz}$) in accordance with a Rabi
oscillations measurement under the same conditions.

The evaluated times of the coherent nuclear spin dynamics in ZnSe:F
under the conditions of the pump-probe experiment are in reasonable
agreement with data for quantum dots, where the carriers are
strongly localized. Our results for the spin dephasing time
$T_{2}^{*}$ and the spin coherence time $T_{2}$ have a similar order
of magnitude as those reported for $\text{GaAs}/(\text{Al,Ga})\text{As}$ QWs~\cite{Sanada2006}
($T_{2}^{*}=90\,\mu\text{s}$ and $T_{2}=270\,\mu\text{s}$ and a
single $\text{GaAs}/(\text{Al,Ga})\text{As}$ QD~\cite{Makhonin2011} ($T_{2}^{*}=16\,\mu\text{s}$ and
$T_{2}=310\,\mu\text{s}$). Note that in
GaAs all nuclei have nonzero spin and the quadrupole interaction
between the nuclei plays an important role in this
system~\cite{Chekhovich2014}.

To summarize, we have demonstrated that the nuclear spin relaxation
processes can be detected all-optically under the conditions of the TRKR
experiment. We employ the advances of optical detection, such as high
sensitivity and spectral selectivity~\cite{Sanada2006}. We
selectively study the spin dynamics of the nuclei in the vicinity of
the fluorine donors, where a spatially inhomogeneous nuclear spin
polarization is established under pulsed laser excitation. The
detection by coherently precessing electrons instead of the
polarization of the luminescence, as commonly used in measurements
of the Hanle effect, allows one to select a single kind of isotope,
since one can work at higher magnetic fields, where the Larmor
precession frequencies of different isotopes split up. Furthermore,
the detection by coherently precessing electrons can be used to
measure the effect of nuclear fields on an ensemble of electron
spins, where one cannot resolve the Zeeman splitting in the nuclear
Overhauser field~\cite{Overhauser1953} spectrally, e.g., as in the
case of single dot spectroscopy~\cite{Makhonin2011,
Chekhovich2013,ChekhovichNP13}.

We study the nuclear spin dynamics of the $^{77}\text{Se}$ isotope
in fluorine-doped ZnSe under the same experimental conditions as in
Ref.~[\onlinecite{Heisterkamp_12_2015}]. The fastest nuclear spin
relaxation time $T_{1}$ or fastest polarization time under these
conditions is found to be in the range from $6$ to $90\,\text{ms}$
at magnetic fields varied from $10$ to $130\,\text{mT}$, while the
longest polarization time is in the range from $100$ to
$430\,\text{ms}$. The nuclear spin coherence time is given by
$T_{2}=520\,\mu\text{s}$, so that the condition $T_{1}\gg T_{2}$ is
valid. Therefore, the spin temperature of the nuclei is established
with the time $T_2$, which occurs much faster than the energy
transfer to the lattice with the $T_1$ time. Thus, the nuclear spin
polarization can be explained using the classical model of nuclear
spin cooling. At this condition the nuclear spin temperature can be
much lower than the lattice temperature.

{\bf Acknowledgments} We acknowledge the financial support by the Deutsche
Forschungsgemeinschaft in the frame of the ICRC TRR 160, the
Volkswagen Stiftung (Project No.\,88360/90080) and the Russian Science
Foundation (Grant No. 14-42-00015). T.K. acknowledges financial
support of the Project SPANGL4Q of the Future and Emerging
Technologies (FET) programme within the Seventh Framework Programme
for Research of the European Commission, under FET-Open Grant No.
FP7-284743. We thank V.~L.~Korenev for helpful discussions.

\end{document}